\documentclass[conference]{IEEEtran}
\IEEEoverridecommandlockouts
\usepackage{cite}
\usepackage{amsmath,amssymb,amsfonts}
\usepackage{algorithmic}
\usepackage{graphicx}
\usepackage{textcomp}
\usepackage{xcolor}
\usepackage{multirow}
\usepackage{array}
\usepackage{url}
\usepackage{booktabs}
\usepackage{verbatim}
\usepackage{diagbox} 

\newcommand{\RNum}[1]{\uppercase\expandafter{\romannumeral #1\relax}}
\newcommand{\tabincell}[2]{\begin{tabular}{@{}#1@{}}#2\end{tabular}}

\def\BibTeX{{\rm B\kern-.05em{\sc i\kern-.025em b}\kern-.08em
    T\kern-.1667em\lower.7ex\hbox{E}\kern-.125emX}}

\begin{document}

\title{Passive TCP Identification for Wired and Wireless Networks:  A Long-Short Term Memory Approach
}

\author{Xiaoyu Chen, Shugong Xu, Xudong Chen, Shan Cao, Shunqing Zhang, Yanzan Sun\\
Shanghai Institute for Advanced Communication and Data Science, \\
Shanghai University, Shanghai, 200444, China\\
Email:\{xiaoyu, shugong, xudongchen, cshan, shunqing, yanzansun\}@shu.edu.cn}

\maketitle

\begin{abstract}

Transmission control protocol (TCP) congestion control
is one of the key techniques to improve network performance. TCP congestion control algorithm identification (TCP identification) can be used to significantly improve network efficiency. Existing TCP identification methods can only be applied to limited number of TCP congestion control algorithms and focus on wired networks. In this paper, we proposed a machine learning based passive TCP identification method for wired and wireless networks. After comparing among three typical machine learning models, we concluded that the 4-layers Long Short Term Memory (LSTM) model achieves the best identification accuracy. Our approach achieves better than 98\% accuracy in wired and wireless networks and works for newly proposed TCP congestion control algorithms.
%
%


\end{abstract}


\section{INTRODUCTION}

%
%

Transmission control protocol / Internet protocol (TCP/IP) lay the foundation of today's information society. With years of research efforts, many TCP algorithms\footnote{For convenience, we will call TCP congestion control algorithm as TCP algorithm in this paper.}  have been devoted to prevent the network congestion as well as the transmission packet losses. As the typical packet transmission over Internet is based on wired architecture, traditional TCP algorithms, such as NewReno\cite{floyd2004newreno}, Cubic\cite{ha2008cubic} and Vegas\cite{brakmo1995tcp}, focus on the network congestion event. With the rapid development of wireless transmission, current TCP algorithms, e.g. Sprout\cite{winstein2013stochastic} / Verus\cite{zaki2015adaptive}, jointly consider the transmission packet loss in the wireless environment, and achieve a better trade-off between the transmission delay and throughput.  

Ideally, the intelligent network shall automatically identify the TCP algorithms and adapt the network routing policy and transmission resources to maximize the network utility. However, even the initial TCP algorithm identification has been regarded as a challenging task, and plenty of research efforts have been spent until recently\cite{yang2014tcp,pahdye2001inferring,medina2005measuring,jaiswal2004inferring,rewaskar2007performance,qian2009tcp}. According to the way of data collection, TCP identification can be mainly divided into two categories, namely active detection and passive measurement, where the former relies on observing the behaviors of injected redundant packets and passive measurement relies on the observations at intermediate nodes, without affecting the current network traffic. The passive measurement has minimum effects on network which is more practical than active detection, so is commonly used in recent work\cite{oshio2009identification,hagos2018general,yang2014tcp}. 

Among the existing passive TCP identification methods, a cluster based scheme has been proposed in\cite{oshio2009identification}, which can identify any 2 out of 14 TCP algorithms and achieves 85\% identification accuracy. However, subject to the cluster-based method, the work is difficult to extend for the identification of newly proposed algorithms. Meanwhile the accuracy is affected by the calculation of artificial features. According to \cite{yang2014tcp}, the congestion avoidance algorithm identification (CAAI) can identify 15 TCP algorithms available in major operating system and reaches overall 96.98\% identification accuracy. The inputs of the model are multiplicative decrease parameter and window growth function (the offset window sizes) that they believe will remain the same for most TCP algorithms. However the assumption is invalid for newly proposed algorithms, like BBR\cite{cardwell2016bbr}, and can not be applied directly. Another kind of passive TCP identification algorithms utilize the back-off factor by inferring the size of congestion window (CWND) during wired  transmissions, which can identify 3 TCP algorithms with 95\% accuracy as reported in \cite{hagos2018general}. Since the back-off factor is time-varying,
this type of method is only valid for loss-based algorithms. As far as we aware, the previous works are based on the wired environment and new types of features need to be exploited for the identification of newly proposed algorithms in wireless networks.

In this paper, we propose a machine learning based passive TCP identification method to address the above issues, and the main contributions are summarized as follows.

\begin{itemize}
     \item{\em Unified Passive TCP Identification Framework.}
     We proposed a unified passive TCP identification framework, which consists of a features extraction block and an identification block. Different from the traditional identification algorithms, the proposed framework can 
     easily expanded to support the identification of new types of TCP algorithms.
     \item{\em Joint consideration of wired and wireless performance.} 
     In order to jointly support wired and wireless environments, we propose a LSTM based passive TCP identification method, which can extract the time-domain correlations to reflect the congestion-based features (mainly for wired networks) and the transmission loss features (mainly for wireless networks). By fully utilizing the LSTM and dense connected neural network architecture, our proposed method can identify 5 TCP algorithms with 99.8\% accuracy in wired networks and 6 TCP algorithms with 98.2\% accuracy in wireless networks.  
     
\end{itemize}

The rest of the paper is arranged as follows. In Section II, the problem formulation and limitations are analyzed. Our system model is introduced in Section III, including features selection and identification model. In Section IV, we introduce the experiment settings and evaluation results. Finally, the conclusion and future work are presented in Section V.

\section{Problem Formulation}

TCP identification methods are typically based on the analysis of maximum a posteriori estimation (MAP), which can be expressed as (\ref{eq8}) according to the Bayes formula. 
\begin{equation}
P_{\theta/F} (\theta_0 /f)=\frac{ P_{F/\theta} (f/\theta_0) * P_\theta (\theta_0)}{P_F (f)}
\label{eq8}
\end{equation}
$\theta$ is the variant of TCP algorithm and $F$ is the input vector. $P_{\theta/F} (\theta_0 /f)$ is the probability of the TCP algorithm $\theta_0$ when the input is the vector $f$. $P_{F/\theta} (f/\theta_0)$ is the probability distribution function of $f$ when TCP algorithm is $\theta_0$. We can further write the above equation (1) in the following form.
\begin{equation}
P_{\theta/F} (\theta_0 /f)=\frac{ P_\theta (\theta_0) }{ P_F (f) } * P_{F/\theta} (f/\theta_0)=C*P_{F/\theta}(f/\theta_0)
\label{eq9}
\end{equation}
$P_{\theta} (\theta_0)$ is the proportion of TCP algorithm $\theta_0$ in the training set. For a Bayesian generative model, the training set and testing set are assumed to be sampled from the same probability distribution\cite{bishop2006pattern}, so $\theta_0$ is a constant here. $P_F (f)$ is the normalization parameter, which represents the distribution probability of the input vector $f$ in the entire input set $F$. As equation (2) is the probability distribution function about TCP $\theta$, the probability of $P_F (f)$ can be considered as a constant. Thus $P_{\theta/F} (\theta_0 /f)$ can be written as the right hand side of equation (2). The problem is transformed to find the maximum probability of all TCP algorithms for each input vector $f$. The TCP algorithm with the maximum probability is considered as the  sender's TCP variant in current input conditions. The optimization function is expressed as shown in (\ref{eq10}).
\begin{equation}
\hat{\theta} = arg \max \limits_{\theta} 
\left\{
{f_{F/\theta}(f/\theta)p_\theta(\theta)}
\right\}
\label{eq10}
\end{equation}
$f_{F/\theta}(f/\theta)$ is the conditional probability density function about input $f$ under TCP algorithm $\theta$. $p_\theta(\theta)$ is the discrete probability mass function of the TCP algorithm. The important part of TCP identification is to construct the probability function.

The previous methods infer the back-off factor $\beta$ for the identification of TCP algorithms\cite{hagos2018general}. For example, the $\beta$ of Cubic is 0.7\cite{ha2008cubic}, for Reno is 0.5. Referring to the traditional Addictive Increase Multiple Decrease (AIMD) mechanism, the core function is (\ref{eq102}).
\begin{equation}
W(t)= \beta (1-p_{t-1})W(t-1)+p_{t-1}[W(t-1)+\alpha]
\label{eq102}
\end{equation}
$W(t)$ is the value of CWND at time $t$. $\alpha$ and $\beta$ are constant. $\alpha$ is used to increase the CWND and $\beta$ is used to decrease the CWND quickly. $p_{t-1}$ indicates a normal ACK when it equals to $1$, otherwise a loss or timeout has occurred in last transmission. 
Two methods are used to infer the $\beta$: (i) using packet loss event; (ii) exploiting time out event. 
By comparing the change in the cwnd value before and after the loss or time-out event occurs, the $\beta$ can be inferred. 
However, even for traditional delay-based algorithms, the above method can not work well. The biggest differences are back-off factors which values change over time. Refer to Vegas \cite{brakmo1995tcp}, the common expressions of $\alpha$ and $\beta$ in delay-based algorithms shown as (\ref{eq106}).
\begin{equation}
\left \{
\begin{aligned}
\alpha_{t} &= m(\frac{TotalSize_{t-1}}{minRTT_{\Delta T}}-\frac{TotalSize_{t-1}}{RTT_{t-1}})^k+C\\
\beta_t &= D(RTT_{t-1},t)
\end{aligned}
\right.
\label{eq106}
\end{equation}
$TotalSize_{t-1}$ is the total bytes sent at last $t$-1 time. $minRtt_{\Delta T}$ is the minimum round trip time (RTT) observed in $\Delta T$ period. $TotalSize_{t-1}/minRtt_{\Delta T}$ stands for the expected transmission rate. $TotalSize_{t-1}/Rtt_{t-1}$ is the actual transmission rate. When actual transmission rate less than expected rate, $\alpha(t) > 0$ and CWND increases, otherwise decreases. $m$, $T$, $k$ and $C$ are constant. $\beta$ is decided by a function about recent observed RTT. More factors, like $RTT$, $minRTT$ and $TotalSize$ which change over time, are introduced to control the CWND. 
According to \cite{yang2014tcp}, the window growth function is introduced to construct the probability function and to realize the identification of the delay-based algorithms.

The newly proposed TCP algorithms that based on different theories and models are more difficult to identify than before. The reason is that they have more control factors that change over time,
and the factors are not independent of each other completely\cite{cardwell2016bbr,zaki2015adaptive,winstein2013stochastic}, which makes the construction of the probability function difficult. The identification model need to deal with multiple control factors so as to identify multiple algorithms. The differences and relations of these factors have to be considered, which is not easy to address in traditional methods. Machine learning has the ability to construct the complex function in this aspect and is used in our method. We will give a further introduction in Sec III.B.  

\section{SYSTEM MODEL}
In this section, the system model of our proposed method  will be introduced, as shown in Fig. 1. In order to extract the features that used to train the network, the captured data passively collected from the user side will be pre-processed firstly. Next step, the pre-processed data is sent to data process block so that it can be aligned on the time-domain. After that the data is sent into the well tuned identification model to identify the TCP algorithm of the data. The parts of features selection and learning-based identification model will be introduced in detail.

\begin{figure*}[!htb] 
\centering 
\includegraphics[height=6cm, width=16cm]{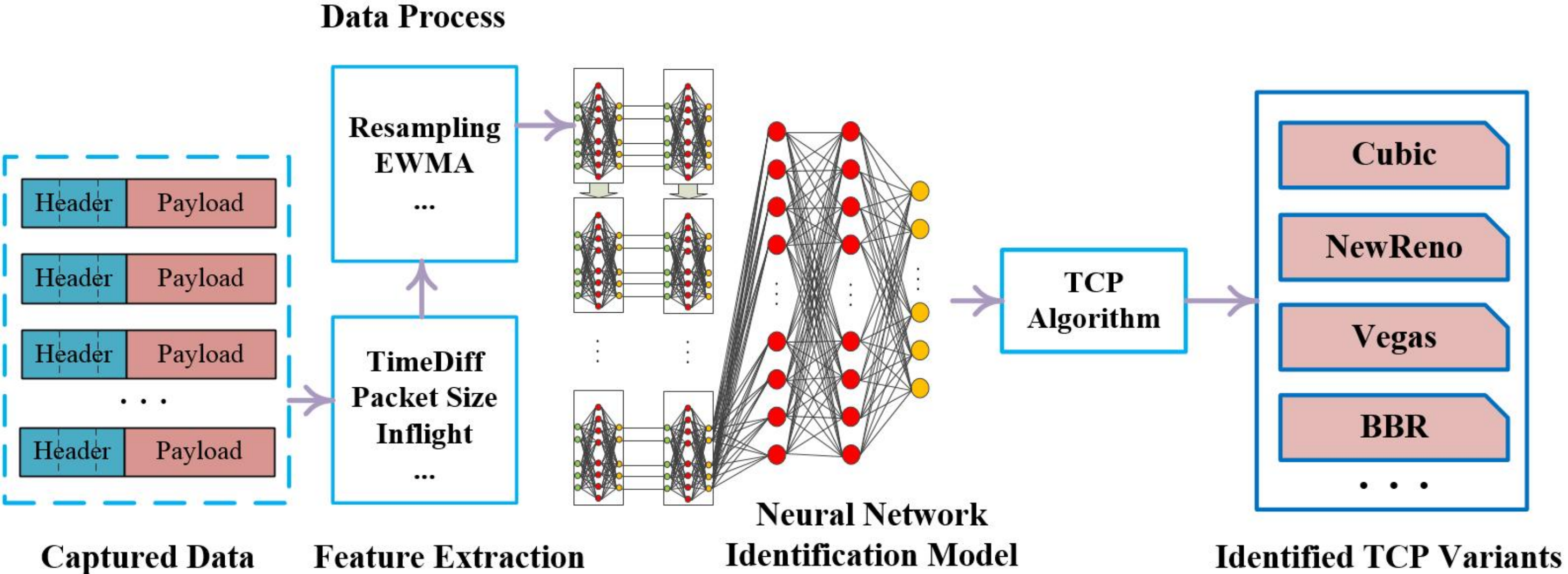} 
\caption{The system model of our proposed method including passive data collection used to extracted  the features and well tuned identification model based on neural network that can accurately identify TCP variants.} 
\end{figure*}

\subsection{Features Selection}

\begin{figure}[h]
\centerline{\includegraphics[height=5cm,width=7cm]{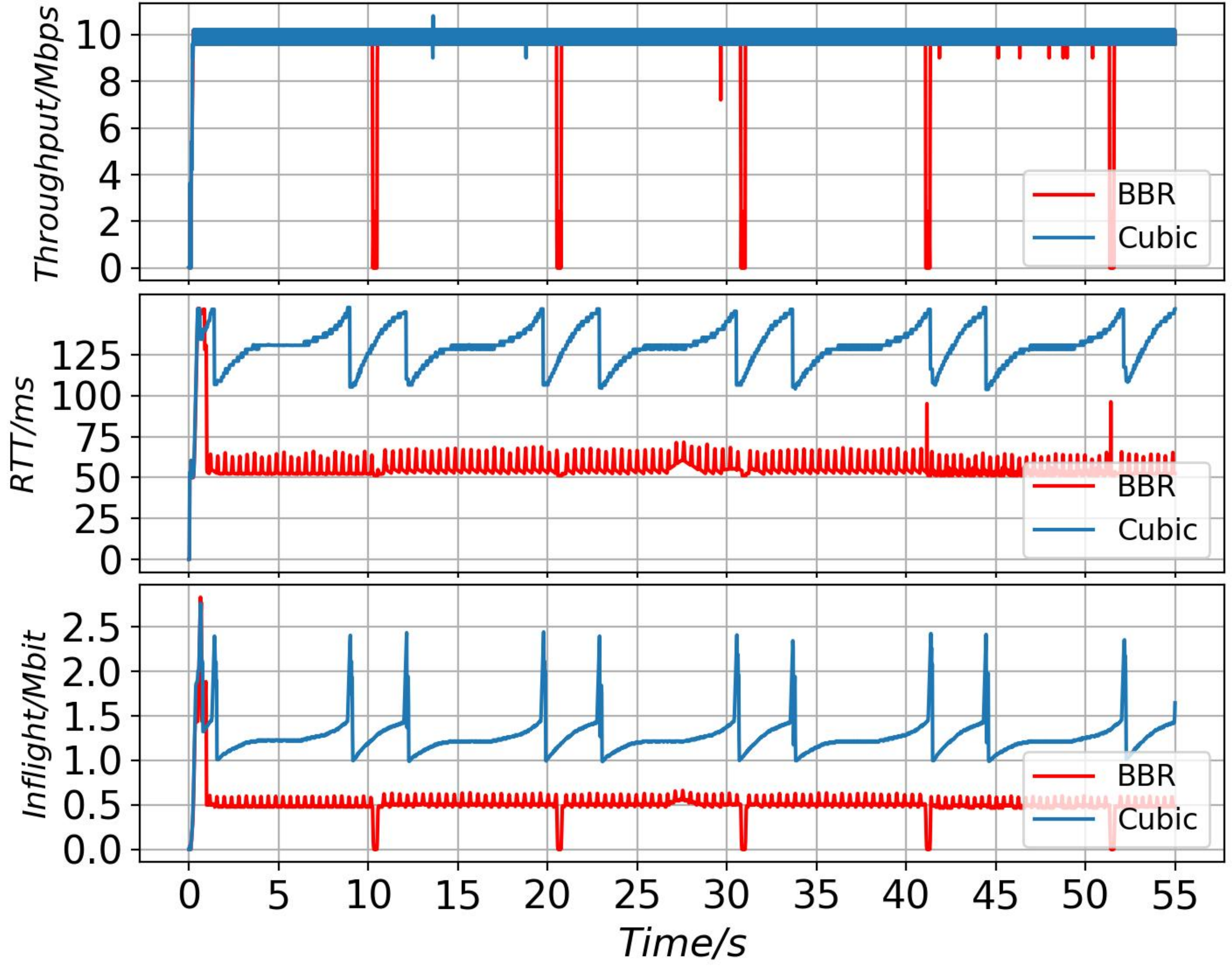}}
\caption{Three features of BBR and Cubic in wired networks. The bottleneck buffer is set to 2BDP, the bandwidth is 10 Mbps and the RTT is 50$ms$.}
\label{fig1}
\end{figure}

\begin{figure}[h] 
\centering 
\includegraphics[height=6cm, width=7cm]{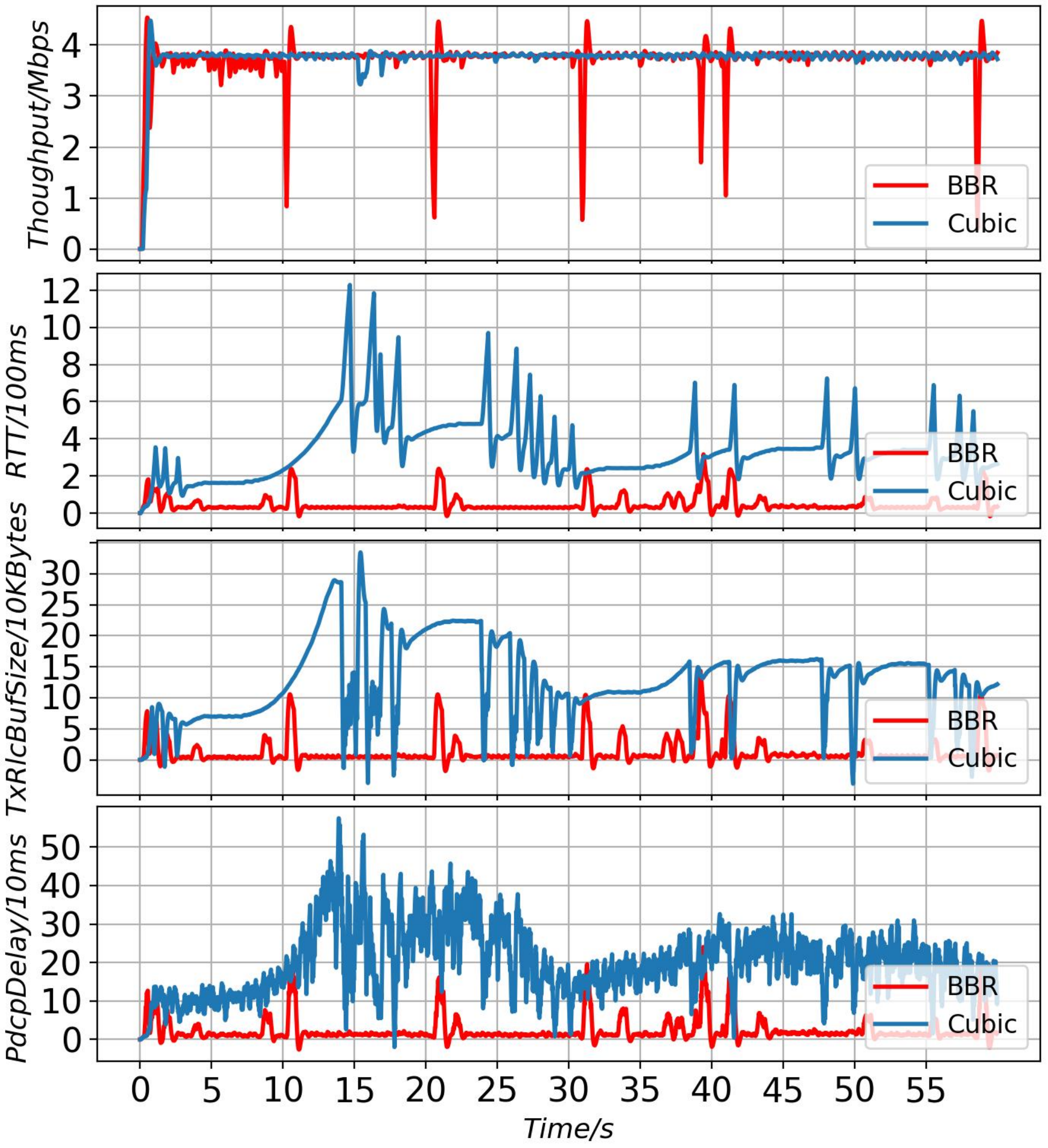} 
\caption{The figure shows the selected features of BBR and Cubic algorithms in wireless networks. The RB numbers is 6, and the RTT sets to 50 $ms$.} 
\end{figure}

For the convenience of actual deployment, the data is collected in a passive way that has the minimal effect on current traffic. Considering different network scenarios, the locations used to collect data differ. The data is collected from the receiver in wired networks, while we choose UE and base station in wireless networks. 

\subsubsection{Features in wired networks}
Although higher accuracy can be achieved if the nodes for data collection are closer to the sender \cite{hagos2018general}. However, server usually runs several service VMs and each of them may use different TCP algorithms which means there are at least two TCP flows in the links from the server to the closest router. One TCP flow may suffer interference from another. We can not change the configuration of router to optimize for a single TCP flow as it is unfair to other flows. In contrast, the reconfiguration of user side is more easy and practical, so the data are collected passively from receiver in wired networks.  

The features extracted from raw TCP packets are passively collected from the receiver side using tcpdump tool, including inflight, throughput and oneway RTT, as shown in Fig. 2. In order to collect oneway RTT, the TCP TimeStamps option is open at the server and user side. And the throughput is calculated as (\ref{eq109}):
\begin{equation}
Throughput = \frac{PacketSize}{TimeDif}
\label{eq109}
\end{equation}
$PacketSize$ is the size of received packet and $TimeDif$ is the difference of arrival time between two adjacent received packets.
The features of different TCP algorithms have its unique behaviors and relationship which will be exploited to train our identification model.

\subsubsection{Features in wireless networks}
Different from wired networks, the state information of base station (BS) is collected besides the raw TCP packets from user equipment (UE) in wireless networks. Because the performance of TCP degrades seriously and the state information from BS can reflect the situation experienced by TCP in wireless channel to some extent. The accuracy of identification will be improved if combined with the state information from BS. 

The features extracted from UE are throughput and oneway RTT\footnote{The inflight has similar behaviors to the changes of RLC buffer size, so it is not considered in wireless networks.}. The state information from BS including the buffer size of radio link control (RLC) layer and the delay of packet data convergence protocol (PDCP) layer. The size of RLC buffer can reflect the overall network congestion. The delay of PDCP layer represents the queuing delay of air interface and the delay caused by re-transmissions in wireless channel, such as automatic repeat request (ARQ) in RLC acknowledged mode (AM).
Compared with the wired networks, wireless channel has high bit errors that result in more fluctuations of the features' behaviors, as shown in Fig. 3, which is a challenge to TCP identification.


\subsection{Learning-Based Identification Model}
\begin{figure}[h] 
\centering 
\includegraphics[height=6.0cm, width=8.5cm]{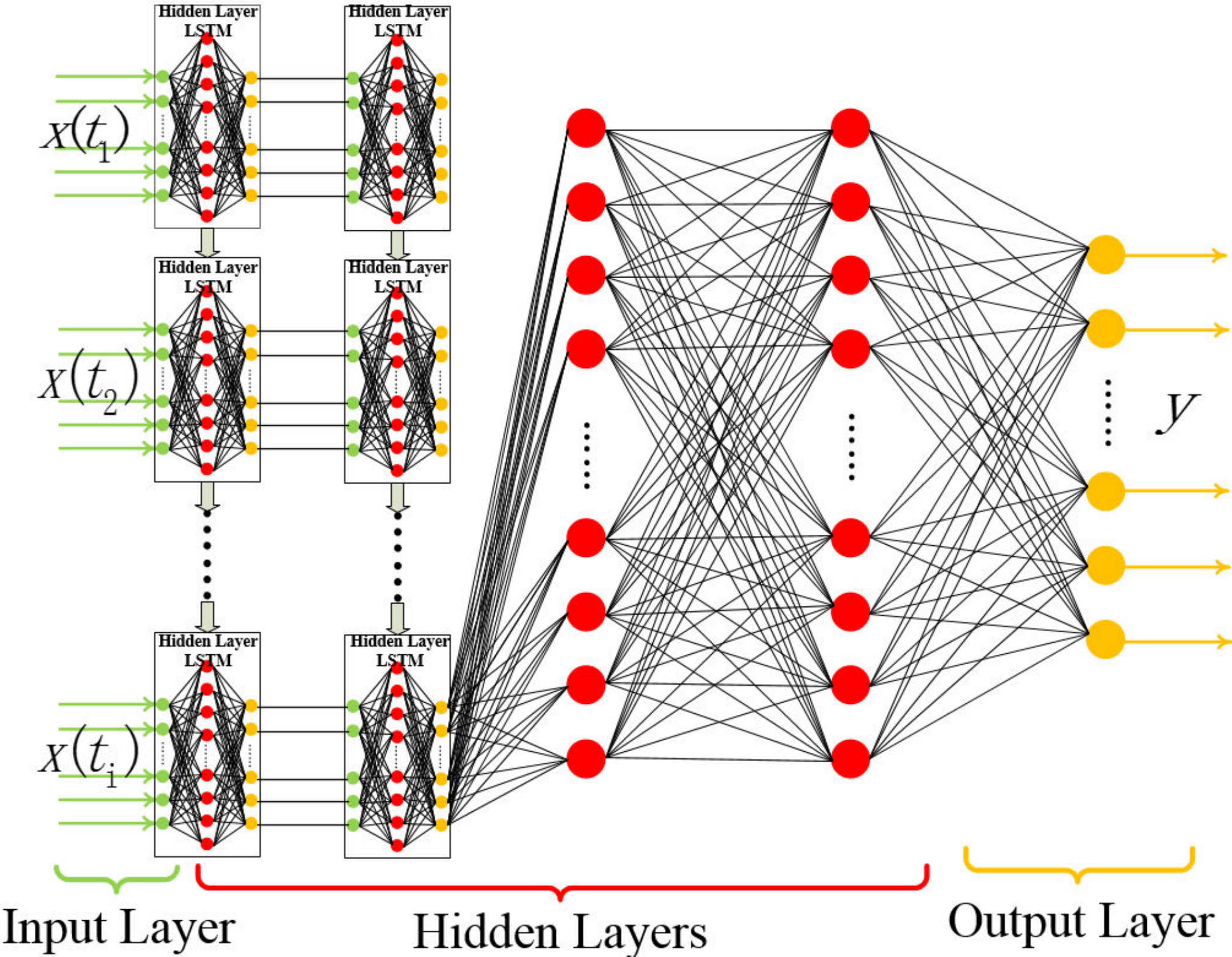} 
\caption{The network model of LSTM} 
\end{figure}

\begin{table}[h]
    \caption{An Overview of Network Configurations and results \label{dnn_cnn_lstm}}
    \centering
    \footnotesize
    \renewcommand\arraystretch{0.5}
    \setlength{\tabcolsep}{2.0mm}{
    \begin{tabular}{c c c c}
        \toprule              &DNN  &CNN  &LSTM   \\
        \midrule
        \midrule  Input Layer &$\mathbb{H}_{3000\times 4}$          &$\mathbb{H}_{60\times 50\times 4}$       &$\mathbb{H}_{20\times 600}$    \\
        \midrule  Layer1      &\tabincell{c}{Dense 1024\\PReLU}     &\tabincell{c}{Conv2D 5x5x128\\s2 PReLU}  &\tabincell{c}{LSTM 600}        \\
        \midrule  Layer2      &\tabincell{c}{Dense 512\\PReLU}      &\tabincell{c}{Conv2D 5x5x64\\ s2 PReLU}  &LSTM 600                       \\
        \midrule  Layer3      &\tabincell{c}{Dense 256\\PReLU}      &\tabincell{c}{Conv2D 3x3x32\\s2 PReLU}   &\tabincell{c}{Dense 256\\PReLU}\\
        \midrule  Layer4      &\tabincell{c}{Dense 128\\PReLU}      &\tabincell{c}{Dense 128\\PReLU}          &\tabincell{c}{Dense 128\\PReLU}\\
        \midrule  Output Layer&\tabincell{c}{Dense 6 \\None softmax}&\tabincell{c}{Dense 6 \\None softmax}    &\tabincell{c}{Dense 6 \\None softmax}\\
        \midrule  \tabincell{c}{Total \\Accuracy}   &0.968 &0.942 &0.982 \\
        \bottomrule
    \end{tabular}}
\end{table}

Machine learning has the ability to extract features and has been widely used in image classification and natural language processing. We combine machine learning with TCP identification as shown in Fig. 1. The data collected from the user side is fed into the neural network to train the machine learning model for the identification of TCP algorithms.

Three typical machine learning models are exploited and compared in wireless networks, including dense neural network (DNN), convolutional neural network (CNN) and LSTM. We use the same dataset to train\footnote{To accelerate the training process of the data set, we install Tensorflow on our server with Intel(R) Xeon(R) CPU E5-3680 and NVIDIA Tesla P100 GPU.} and test, and all network models have been well tuned to achieve the best accuracy. The network configurations and results are listed in Table I. The 4-layers LSTM network has the highest identification accuracy, and its model shown in Fig. 4.

The reason for poor performance of DNN is that the output of fully connected DNN is only decided by current state which means the timing information is not considered. Although CNN has strong ability to extract features, the features of different time are not considered either. The LSTM model that combines the information from the current state with the previous state to get the output is suited for identifying because it takes account of the time series. In our well tuned model the number of previous state which we consider is 131. The model also has high scalability, when a new TCP algorithm join the model, the model can be fine-tuned using most of existed parameters except for the output softmax layer, which means the model will converge quickly.
The training property of the LSTM used in this paper will be introduced, including the input and output, training dataset, data process, loss function, and other implementation details.

{\bfseries Input and output.}
The input data is a set of features with time series which is different between wired and wireless networks. In wired networks, the input consists of PacketSize, TimeDif, OnewayDelay and Inflight. In wireless networks, PacketSize, TimeDif, OnewayDelay, PDCPDelay, and RLCBufferSize are included. The output is the TCP algorithm variant of the input data. 

{\bfseries Training dataset.}
The dataset is generated using mininet and ns-3 which will be introduced in Sec IV.A. We use 16800 sets of data to train wired identification model and 4320 sets of data to train wireless model.

{\bfseries Data process.}
The data directly obtained from the networks can not be aligned on the time-domain, which will affect the training and testing of LSTM model. We re-sample the data using linear interpolation and the sampling interval is set to 5 $ms$. The exponential weighted moving average method is then used to reduce the impact of interpolation operation and the discontinuity of input data.


{\bfseries Loss function.}
The loss function we use is cross entropy calculated as equation (\ref{eq7}) shows. $x$ is the input data of the network and $c$ is the label.

\begin{equation}
\left \{
\begin{aligned}
L_{conf}(x,c) &= - \sum_{i\in Pos}^{N}{x_{ij}^{p}}log(\hat{c}_{i}^p)- \sum_{i \in Neg} log(\hat{c}_{i}^0) \\
\hat{c}_i^p &= \frac{exp(c_i^p)}{\sum_p exp(c_i^p)}
\end{aligned}
\right.
\label{eq7}
\end{equation} 

{\bfseries Other implementation details.}
The total epochs used to train the LSTM model is 500. The batch size is 32. We use Adam optimizer with $10^{-4}$ initial learning rate and reduce the learning rate by ten times when in $0.5$ and $0.7$ total epochs, respectively.


\section{Experiments and Evaluations}
In this section, the experiment settings of wired and wireless networks topology and evaluation results are given in detail.

\begin{figure}[h] 
\centering 
\includegraphics[height=3.5cm, width=8.5cm]{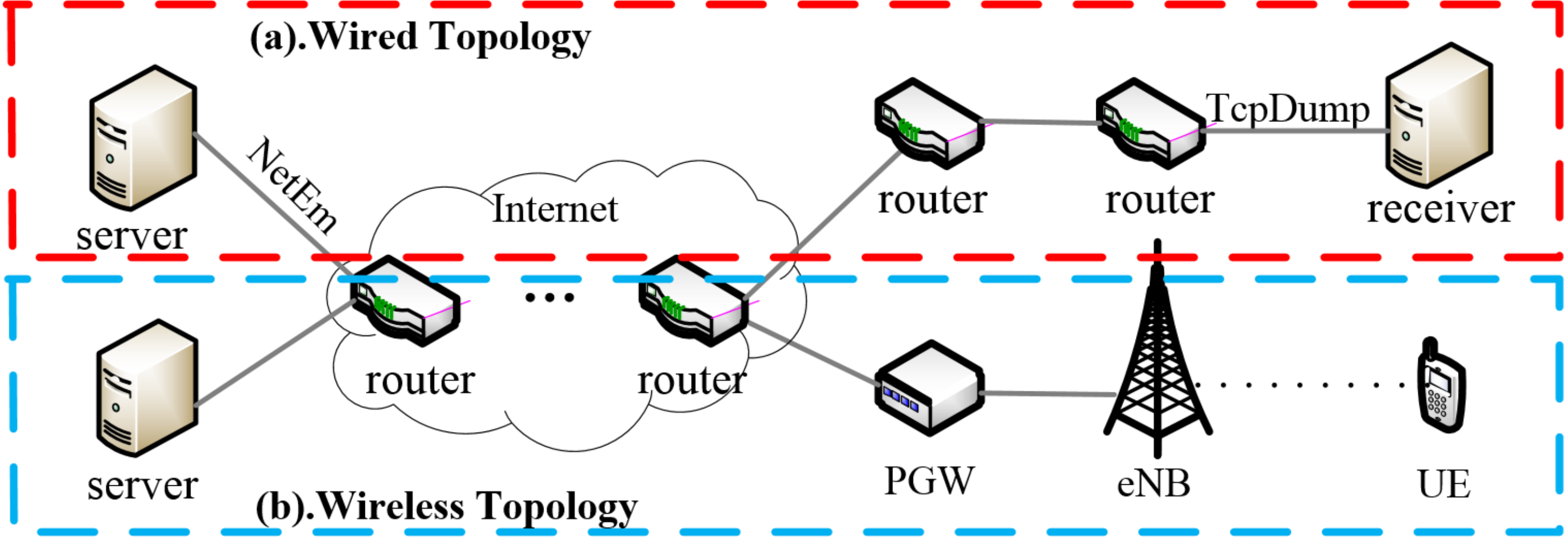} 
\caption{The topology of wired and wireless networks.} 
\end{figure}

\subsection{Networks Topology and Parameters Setup}

\subsubsection{\bfseries{Wired Network Topology}}
The wired network topology is constructed using mininet\cite{mininetoline} as shown in Fig. 5 (a). The server connects to a router and Netem is used to set links delay from the server to the receiver. The rate limiting and buffer size at the router is configured by Token-Bucket Filter\cite{scholztowards}. Tcpdump is used to passively collect data from the receiver.

The TCP algorithms used to generate the dataset for training including Cubic, NewReno, Hybla, Vegas and BBR, which are commonly used in current networks\footnote{The Linux kernel used in simulation is 4.13, which contains the TCP congestion control algorithms we need.}. 
The maximum bandwidth is 5$\sim$10 $Mbps$, and the range of link RTT is from 40 to 100 $ms$. The size of bandwidth delay product (BDP) is 200$\sim$1000 packets\footnote{The maximum transmission unit sets to 1500 Bytes.}. We repeat each simulation 10 times for 60 $s$ and generated 16,800 sets of data for training, and 900 sets of data are generated for testing. 

\subsubsection{\bfseries{Wireless Network Topology}}
The ns-3\cite{ns-3Online} is used to construct the wireless network topology as shown in Fig. 5 (b). Compared with wired network, the Westwood is added because of its good performance in wireless channel. The implementations of Cubic\cite{levasseur2014tcp} and BBR\cite{claypool2018bbr} have been added to the original ns-3. The RLC AM mode is used in actual deployment, so is also considered in simulation. The maximum size of RLC buffer is unlimited in original ns-3, we modified the implementation of AM to support the drop-tail mechanism\cite{robert2016behaviour}. The tracing system of ns-3 is turned on to collect the TCP packets from receiver side. As for the delay of PDCP layer and RLC buffer size, which can be extracted from log files.

In terms of bandwidth, the number of resource block sets to 6, 15 and 25, respectively. 
The RTT is 20$\sim$100 $ms$ and the delay of the link between server and the closest router is adjust accordingly. The delay of air interface between BS and UE is considered as 3 $ms$. The maximum RLC buffer size sets to 100$\sim$700 packets. The error model is MiErrorModel and the fading model is EVA60kmph. The Bulksend is used to generate data at the server side. Finally, 4320 sets of data are generated for training and 1440 sets of data for testing.

\subsection{Performance Evaluations}

\begin{table}[t]
    \caption{The confusion matrix in wired networks \label{wired_result1}}
    \centering
    \footnotesize
    \renewcommand\arraystretch{0.3}
    \setlength{\tabcolsep}{3.0mm}{
    \begin{tabular}{c c c c c c}
        \toprule  & BBR &Cubic &NewReno &Hybla &Vegas\\
        \midrule 
        \midrule BBR               & 179 & 1   & 0   & 0   & 0\\
        \midrule Cubic             & 0   & 179 & 1   & 0   & 0\\
        \midrule NewReno           & 0   & 0   & 180 & 0   & 0\\
        \midrule Hybla             & 0   & 0   & 0   & 180 & 0\\
        \midrule Vegas             & 0   & 0   & 0   & 0   & 180\\
        \bottomrule
    \end{tabular}}
\end{table}

\begin{table}[h]
    \caption{The identification results in wired networks \label{wired_result2}}
    \centering
    \footnotesize
    \renewcommand\arraystretch{0.3}
    \setlength{\tabcolsep}{3.0mm}{
    \begin{tabular}{c c c c c}
        \toprule   & \textit{Precision} & \textit{Recall} & \textit{F1-Score} & \textit{Support}  \\
        \midrule 
        \midrule BBR               & 1.000  &0.994  &0.998  &180\\
        \midrule Cubic             & 0.994  &0.994  &0.994  &180\\
        \midrule NewReno           & 0.995  &1.000  &0.997  &180\\
        \midrule Hybla             & 1.000  &1.000  &1.000  &180\\
        \midrule Vegas             & 1.000  &1.000  &1.000  &180\\
        \midrule Average/Total     & 0.998  &0.998  &0.998  &900\\
        \midrule Accuracy          & 0.998 \\
        \bottomrule
    \end{tabular}}
\end{table}

\begin{table}[!htb]
    \caption{The confusion matrix in wireless networks \label{wireless_result1}}
    \centering
    \footnotesize
    \renewcommand\arraystretch{0.3}
    \setlength{\tabcolsep}{1.5mm}{
    \begin{tabular}{c c c c c c c}
        \toprule   & BBR &Cubic &NewReno &Hybla &Vegas &Westwood\\
        \midrule 
        \midrule BBR        &240 &0   &0   &0   &0   &0  \\
        \midrule Cubic      &0   &227 &13  &0   &0   &0  \\
        \midrule NewReno    &0   &0   &238 &0   &2   &0  \\
        \midrule Hybla      &0   &2   &9   &229 &0   &0  \\
        \midrule Vegas      &0   &0   &0   &0   &240 &0  \\
        \midrule Westwood   &0   &0   &0   &0   &0   &240\\
        \bottomrule
    \end{tabular}}
\end{table}

\begin{table}[!htb]
    \caption{The identification results in wireless networks \label{wireless_result2}}
    \centering
    \footnotesize
    \renewcommand\arraystretch{0.3}
    \setlength{\tabcolsep}{3.0mm}{
    \begin{tabular}{c c c c c}
        \toprule   & \textit{Precision} & \textit{Recall} & \textit{F1-Score} & \textit{Support}  \\
        \midrule 
        \midrule BBR               & 1.000  &1.000  &1.000  &240\\
        \midrule Cubic             & 0.991  &0.946  &0.968  &240\\
        \midrule NewReno           & 0.915  &0.992  &0.952  &240\\
        \midrule Hybla             & 1.000  &0.954  &0.976  &240\\
        \midrule Vegas             & 0.992  &1.000  &0.996  &240\\
        \midrule Westwood          & 1.000  &1.000  &1.000  &240\\
        \midrule Average/Total     & 0.983  &0.982  &0.982  &1440\\
        \midrule Accuracy          & 0.982 \\
        \bottomrule
    \end{tabular}}
\end{table}

The identification results of our proposed method are analyzed in this section. In terms of accuracy, we mainly use the confusion matrix and the criterias of precision, recall and F1-score. The identification accuracy in wired networks is listed in Table II and Table III, from which we can see that almost all TCP algorithms are accurately identified and the overall accuracy is 99.8\%.

The model shows a high accuracy on the identification of BBR, Vegas and Westwood in wireless networks from Table IV and Table V. The identification accuracy of the other three algorithms is no less than 94.6\%, which also demonstrates that our proposed method performs well in wireless networks.

\begin{figure}[h] 
\centering 
\includegraphics[height=8 cm, width=8cm]{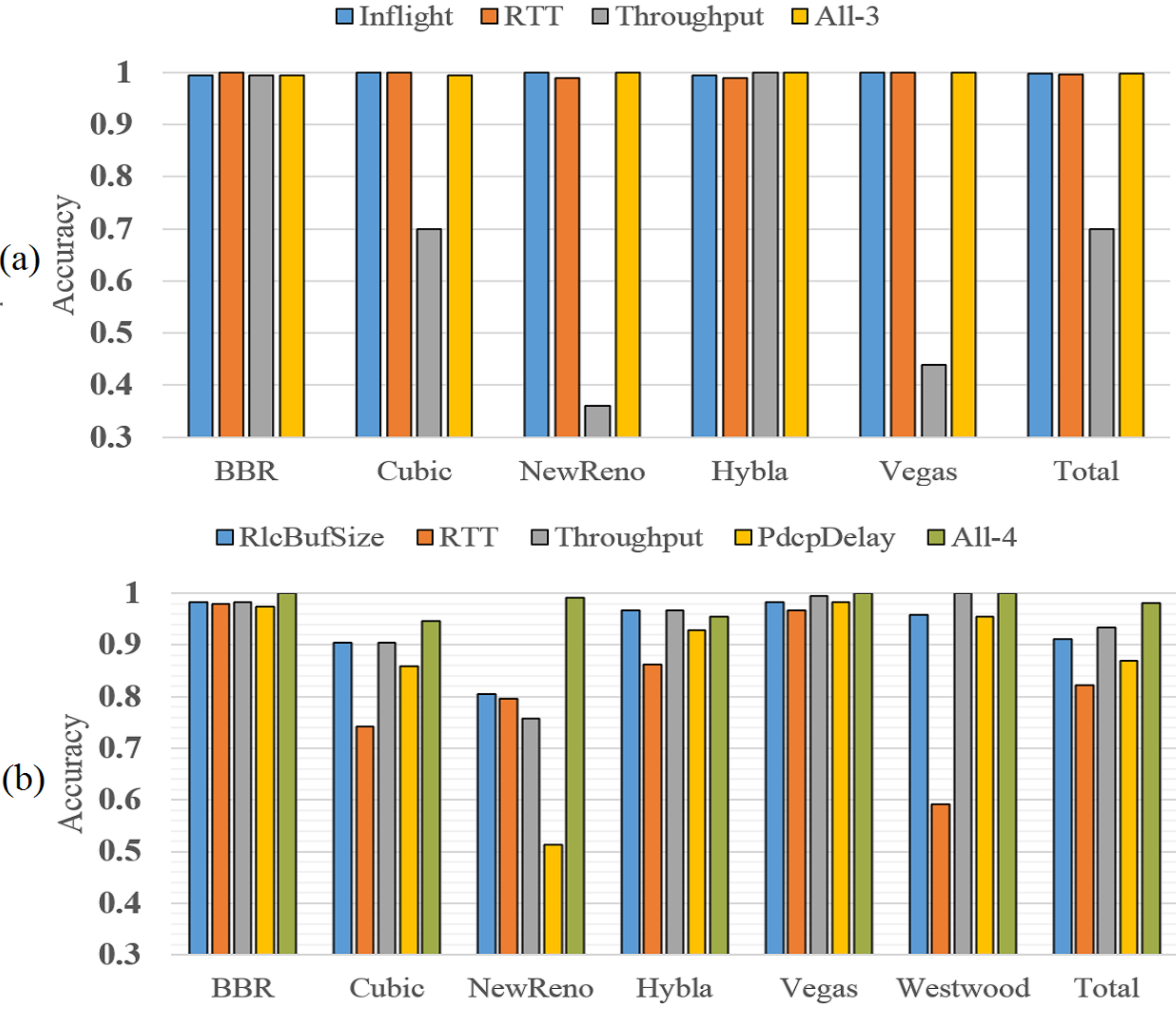} 
\caption{The impact of selected features on identification accuracy. 
(a) shows the impact of input features on the accuracy of each TCP algorithm in wired networks and (b) shows their impact in wireless LTE networks.} 
\end{figure}

The impact of selected features on the accuracy of each TCP algorithm identification is analyzed, as shown in Fig. 6, which provides a reference for selecting appropriate input for future optimization. Fig. 6 (a) shows that all features have high accuracy on BBR. However the performance is different on the other algorithms, especially for throughput. The reason is the behavior of throughput is steady if no loss occurs when use loss-based algorithms. Fig. 6 (b) shows that the selected features have large fluctuations on accuracy although the overall accuracy is above 94.6\%, which is caused by the error-prone wireless channel. The accuracy of throughput is higher in wireless networks than in wired networks. In contrast, the RTT has a poor performance than in wired networks.

\section{CONCLUSION AND FUTURE WORK}
In this paper, a machine learning based passive TCP identification method is proposed. The proposed method can realize high identification accuracy for loss based, delay based and even the newly proposed algorithms with high scalability. Our method can be used in both wired and wireless network that covers most of scenes in current network. More than 98.2\% identification accuracy is achieved in both network scenarios. 

This work will be extended in the following aspects.
1) combine the identification method with actual system to test its performance, including optimizing configurations after identifying the variant of TCP algorithms; 2) update the identification model under multiple flows competitions.

\section*{Acknowledgment}
This work was supported by the National Natural Science Foundation of China (NSFC) Grants under No. 61701293 and No. 61871262, the National Science and Technology Major Project Grants under No. 2018ZX03001009, the Huawei Innovation Research Program (HIRP), and research funds from Shanghai Institute for Advanced Communication and Data Science (SICS).

\bibliography{reference}{}
\bibliographystyle{IEEEtran}
\end{document}